\documentclass[12pt,a4paper]{article}
\pdfoutput=1
\usepackage{graphicx,amssymb,amstext,amsmath,cases,float,setspace,mathtools,amsthm,physymb,cite}
\usepackage[margin=1.0in]{geometry}
\begin{document}
\title{Dimensional Transitions in Thermodynamic Properties of Ideal Maxwell-Boltzmann Gases\thanks{NOTICE: This is an author-created, un-copyedited version of an article accepted for publication in Physica Scripta. IOP Publishing Ltd is not responsible for any errors or omissions in this version of the manuscript or any version derived from it. The Version of Record is available online at doi:10.1088/0031-8949/90/4/045208.}}
\author{Alhun Aydin and Altug Sisman\thanks{Corresponding author, e-mail: sismanal@itu.edu.tr}}
\maketitle
\begin{center}
\small\textit{Istanbul Technical University, Energy Institute, 34469, Istanbul, Turkey.}
\end{center}
\begin{abstract}
An ideal Maxwell-Boltzmann gas confined in various rectangular nano domains is considered under quantum size effects. Thermodynamic quantities are calculated from their relations with partition function which consists of triple infinite summations over momentum states in each direction. To get analytical expressions, summations are converted to integrals for macro systems by continuum approximation which fails at nanoscales. To avoid both from the numerical calculation of summations and the failure of their integral approximations at nanoscale, a method which gives an analytical expression for single particle partition function (SPPF) is proposed. It's shown that dimensional transition in momentum space occurs at certain magnitude of confinement. Therefore, to represent SPPF by lower-dimensional analytical expressions becomes possible rather than numerical calculation of summations. Considering rectangular domains with different aspect ratios, comparison of the results of derived expressions with those of summation forms of SPPF is done. It’s shown that analytical expressions for SPPF give very precise results with maximum relative errors of around $1\%$, $2\%$ and $3\%$ at just the transition point for single, double and triple transitions respectively. Based on dimensional transitions, expressions for free energy, entropy, internal energy, chemical potential, heat capacity and pressure are given analytically valid for any scale.
\vspace{2.5 mm}
\end{abstract}
\section{Introduction}
Advancements in nanoscience and nanotechnologies increased the importance of studies related to the nanoscale properties of matter, in recent years. On the presence of quantum confinement effects, thermodynamic properties of systems differ from those of macro scale. Due to wave character of particles, quantum size effects (QSE) appear at nanoscale systems as a consequence of quantum confinement.

Thermodynamics under QSE became an active research area and many novel aspects such as quantum boundary layer, anisotropic gas pressure, loss of additivity in extensive properties, quantum forces and gas diffusion due to size difference in both Maxwell-Boltzmann (MB) and quantum (Fermi-Dirac and Bose-Einstein) gases have been studied in literature\cite{nature,molina,pathria,dai1,dai2,naturem,sismanmuller,sisman,dai3,qbl,uqbl,nanocav,karabulut,pdx,babac,qforce,aydin}. Even though size and shape dependencies in thermodynamics of nano systems has been recently studied, dimensional transitions due to QSE in thermodynamic quantities and transition conditions have not been examined in literature.

Thermodynamic state functions are represented by summations from their definitions in statistical mechanics\cite{kittel,pathbook}. In thermodynamics of macro systems, it is common to replace summations by integrals to decrease the calculation burden and to obtain analytical expressions. Conversion of summations to integrals is done by using continuum approximation and density of states concept, under the assumption of discrete energy levels are so close to each other that they can be considered as continuous variables. Continuum approximation works extremely well in macro scale. However, in nanoscale, ascending divergence appears between summations and their integral approximations\cite{sismanmuller,pathbook}.

At meso scales, instead of integral approximations, Weyl's conjecture or first two terms of Poisson summation formula (PSF) are used in literature\cite{molina,pathria,dai1,dai2,sismanmuller,sisman,dai3,qbl,uqbl,nanocav,pathbook}. On the other hand, when characteristic sizes of the domain are smaller than half of the most probable de Broglie wavelengths of particles, even these approximations considered in literature become insufficient. In the case of such strong confinements, one should either consider the numerical calculation of exact summations in thermodynamic properties or look for a better solution.

In this article, a method based on dimensional transitions in momentum space due to quantum confinement is proposed for the calculation of single particle partition function (SPPF) of MB statistics. It is seen that the method leads to analytical expressions which give nearly the same results of exact summations. Then from the analytical expressions of SPPF, thermodynamic quantities such as free energy, entropy, internal energy, heat capacity and pressure are derived analytically. Consequently, the results successfully represent thermodynamic behavior of strongly confined systems without need to calculate relevant triple summations.

\section{Determination of Dimensional Transition Point in Momentum Space}
For a high temperature and/or low density ideal gas, SPPF is given in triple summation form by using MB statistics as follows
\begin{equation}
\zeta=\sum_{i_n}\exp{\left(-\frac{\varepsilon_{i_n}}{k_B T}\right)}
\end{equation}
where $i_n=\left\{1,2,3,...\right\}$ are momentum state variables with subscript of $n=\left\{1,2,3\right\}$ indicating directions, $\varepsilon$ are the energy eigenvalues from the solution of the Schr\"odinger equation for a rectangular domain, $k_B$ is Boltzmann's constant and $T$ is temperature. By defining a confinement parameter $\alpha_n=L_c/L_n$ that indicates the magnitude of confinement in $n$'th direction, Eq. (1) can also be written as follows
\begin{equation}
\zeta=\prod_{n=1}^{3}\sum_{i_n=1}^{\infty}{\exp(-\alpha_n i_n)^2}
\end{equation}
In the definition of $\alpha$, $L_n$ is the size of the domain in the $n$'th direction and $L_c=h/\sqrt{8mk_B T}$ is the half of the most probable de Broglie wavelength of particles where $h$ is Planck's constant and $m$ is mass of the particle.

Although triple summations are the exact way to calculate SPPF, in macro scales where $\alpha$ values are much smaller than unity ($\alpha<<1$), summations are replaced by integrals making continuum approximation. On the other hand, when domain size in particular direction ($L_n$) is on the order of $L_c$, like in nanoscale ($\alpha\approx 1$), there is a considerable deviation between the results of summations and integrals. Therefore, a more precise evaluation method, such as Poisson summation formula (PSF), is necessary to calculate summations. PSF is written for even functions, $f(i)=f(-i)$, as
\begin{equation}
\sum^{\infty}_{i=1} f(i)=\int^{\infty}_{0} f(i)di - \frac{f(0)}{2}+2\sum^{\infty}_{s=1}\int^{\infty}_{0} f(i)cos(2\pi si)di
\end{equation}
By evaluating each term of PSF, single particle partition function in 1D can be exactly represented as
\begin{subequations}
\begin{align}
\zeta_{1D} &=\sum^{\infty}_{i_1=1}\exp\left[{-(\alpha_1 i_1)^2}\right]\\
&=\frac{\sqrt{\pi}}{2\alpha_1}-\frac{1}{2}+\frac{\sqrt{\pi}}{\alpha_1}\sum^{\infty}_{s=1}\exp\left[{-\left(\frac{\pi s}{\alpha_1}\right)^2}\right]
\end{align}
\end{subequations}

The first term of PSF is the conventional integral term and it's accuracy for macro domains $(\alpha\rightarrow 0)$ is quite well. When domain size gets closer to $L_c$ $(\alpha\rightarrow 1)$, contribution of the second term (called zero correction term) of PSF becomes appreciable. First two terms of PSF can be called one-dimensional (1D) representation, since it represents the contributions of whole momentum state variables, $i_1=\left\{1,...,\infty\right\}$. Contribution of the zero correction term is studied as QSE in literature\cite{sismanmuller,sisman,qbl,uqbl,nanocav,babac,qforce} especially for weakly confined domains $(\alpha<1)$ in which the contribution of the third term of PSF is negligible. To fully represent SPPF in any scale, even for strongly confined domains $(\alpha>1)$, whole terms of PSF have to be considered. However, since there is no known analytical solution for the third term which include another infinite summation, either we have to calculate infinite summations of SPPF numerically or find a method to give analytical expressions representing the exact behavior of SPPF successfully. For strongly confined cases $(\alpha>1)$, fortunately, contribution of just the first term ($i_1=1$) of SPPF's summation becomes extremely dominant. Therefore, it can be called as zero-dimensional (0D) representation since it represents almost only the contribution of the ground state.

To see the success of 1D and 0D representations, variations of their analytical expressions with confinement parameter are compared with that of exact 1D SPPF based on numerical calculation of summation, in Figure 1. The solid curve shows the variation of exact form of 1D SPPF (Eq. 4a) with confinement parameter $\alpha_1$. Plus signs show the results of 1D analytical representation based on first two terms of PSF, while cross signs demonstrate the results of 0D representation based on just the first term ($i_1=1$) of SPPF's summation.

\begin{figure}[H]
\centering
\includegraphics[width=0.75\textwidth]{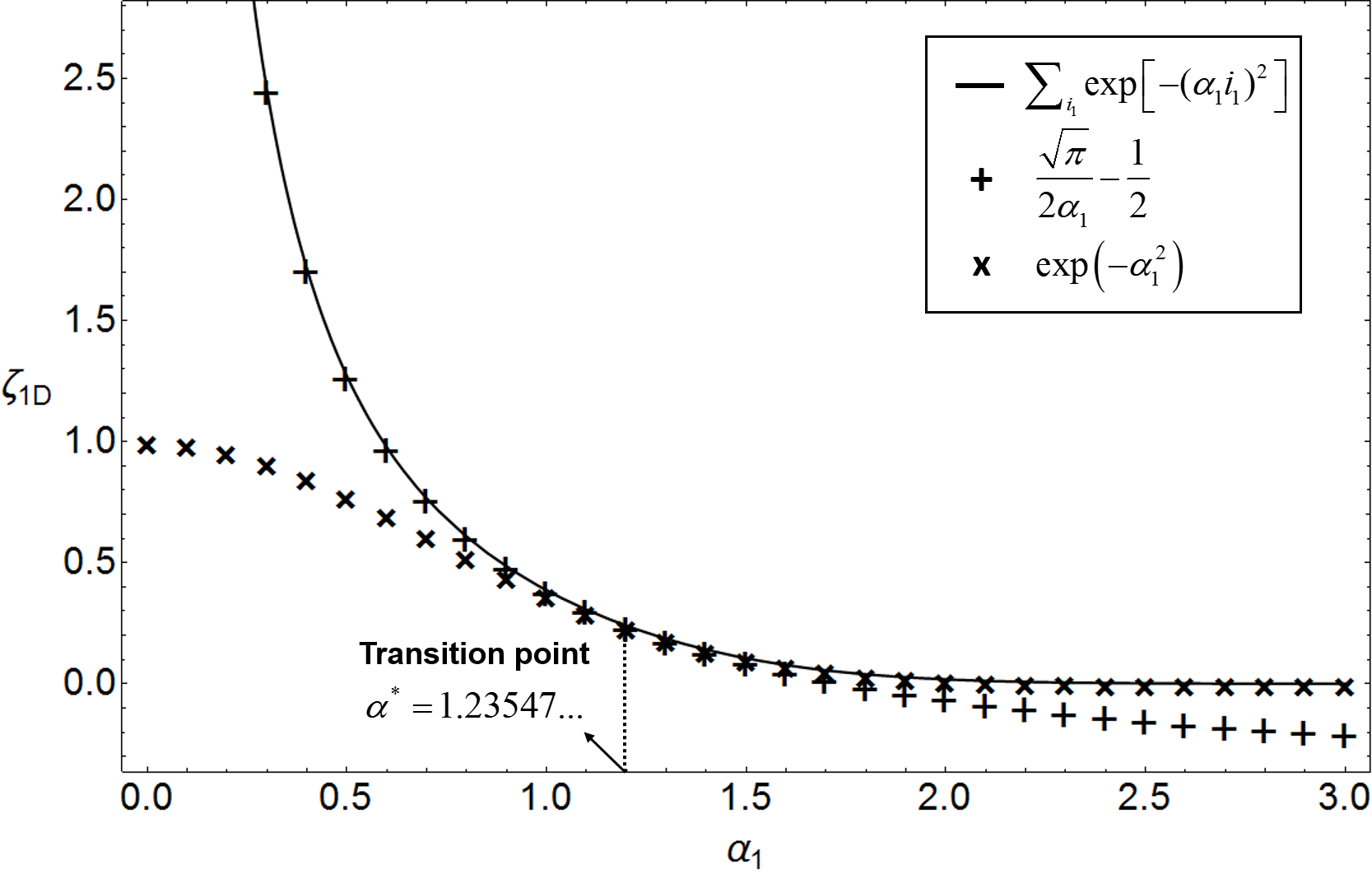}
\caption{Dimensional transition from 1D to 0D representation in SPPF for an MB gas}
\label{fig:pic1}
\end{figure}

It is understood that 1D analytical expression matches perfectly with the exact form of SPPF until the certain value of $\alpha=\alpha^{*}$, after that it considerably deviates. In the same manner, 0D expression substantially deviates before $\alpha^{*}$, while it perfectly matches with the solid curve after that. This certain value of $\alpha$, which corresponds to the transition point from 1D to 0D representation, can be found as $\alpha^{*}=1.23547...$ by equating 1D expression to 0D one. Therefore, an analytical expression representing SPPF for any values of $\alpha$ can be given as follows
\begin{equation}
{\sum^{\infty}_{i_1=1}\exp\left[{-(\alpha_1 i_1)^2}\right]\approx\left(\frac{\sqrt{\pi}}{2\alpha_1}\right)^c\left[\exp({-\alpha_1^2})\right]^{1-c}-\frac{c}{2}}
\begin{cases}
	c=1   & \text{for}  \;\;  \alpha_1 < \alpha^{*} \\
	c=0   & \text{for}  \;\;  \alpha_1 \geq \alpha^{*}
\end{cases}
\end{equation}
where the transition value is $\alpha^{*}=1.23547...$. This approximation gives $1\%$ error even at the transition point. The transition represents the dimensional change (contraction/expansion) of momentum space in particular direction. Henceforward, the domain in the $n$'th direction will be considered as free and confined for $\alpha_n < \alpha^{*}$ and $\alpha_n \geq \alpha^{*}$ respectively.

\subsection{Dimensional Transitions and Superpositions in Single Particle Partition Function}

By using the dimensional transition point in momentum space, we may now consider SPPFs for the domains which are confined at nanoscale in certain directions. 3D domain is free in all three directions, so that confinement values in all directions are smaller than $\alpha^{*}$, the transition value. For 2D domain, $\alpha_1$ and for 1D domain, $\alpha_1$ and $\alpha_2$ are chosen as greater than $\alpha^{*}$ respectively. 0D domain is considered as confined in all directions, which means all of their confinement parameters are larger than $\alpha^{*}$. In this sense, using Eq. (5), we can write analytical expressions for SPPF of an ideal MB gas for 3D, 2D, 1D and 0D domains as

\begin{equation}
\zeta_a=
\begin{dcases}
\zeta_{3D}&=\frac{\pi^{3/2}}{8\alpha_{1}\alpha_{2}\alpha_{3}}\left(1-\frac{\alpha_{1}}{\sqrt{\pi}}\right)\left(1-\frac{\alpha_{2}}{\sqrt{\pi}}\right)\left(1-\frac{\alpha_{3}}{\sqrt{\pi}}\right) \\
\zeta_{2D}&=\frac{\pi}{4\alpha_{2}\alpha_{3}}\exp\left(-\alpha_{1}^2\right)\left(1-\frac{\alpha_{2}}{\sqrt{\pi}}\right)\left(1-\frac{\alpha_{3}}{\sqrt{\pi}}\right), \quad \{\alpha_1\}>\alpha^{*} \\
\zeta_{1D}&=\frac{\sqrt{\pi}}{2\alpha_{3}}\exp\left(-\alpha_{1}^2-\alpha_{2}^2\right)\left(1-\frac{\alpha_{3}}{\sqrt{\pi}}\right), \quad \{\alpha_1,\alpha_2\}>\alpha^{*} \\
\zeta_{0D}&=\exp\left(-\alpha_{1}^2-\alpha_{2}^2-\alpha_{3}^2\right), \quad \{\alpha_1,\alpha_2,\alpha_3\}>\alpha^{*}
\end{dcases}
\end{equation}

Although dimensional transition point has a unique value, by engaging the sizes of the domain to each other in each direction, it is possible to obtain different transition points correspond to different dimensional transitions for anisometric domains. In order to engage sizes to each other, we may define constant aspect ratios between sizes in three directions. Therefore, we can change the size of the domain by changing only one of the sizes, which is chosen here as $L_1$, while preserving its shape. Aspect ratio between sizes in first and second directions is denoted by $r_{12}=L_1/L_2$ and while the second and third directions by $r_{23}=L_2/L_3$. In that case, confinement parameters of second and third directions can be written in terms of the confinement parameter in the first direction as $\alpha_2=\alpha_1 r_{12}$ and $\alpha_3=\alpha_1 r_{12}r_{23}$ respectively. In Figure 2, relative errors ($R_{\zeta_a}$) of analytical expressions given by Eq. (6) are given for different aspect ratios.

\begin{figure}[H]
\centering
\includegraphics[width=1\textwidth]{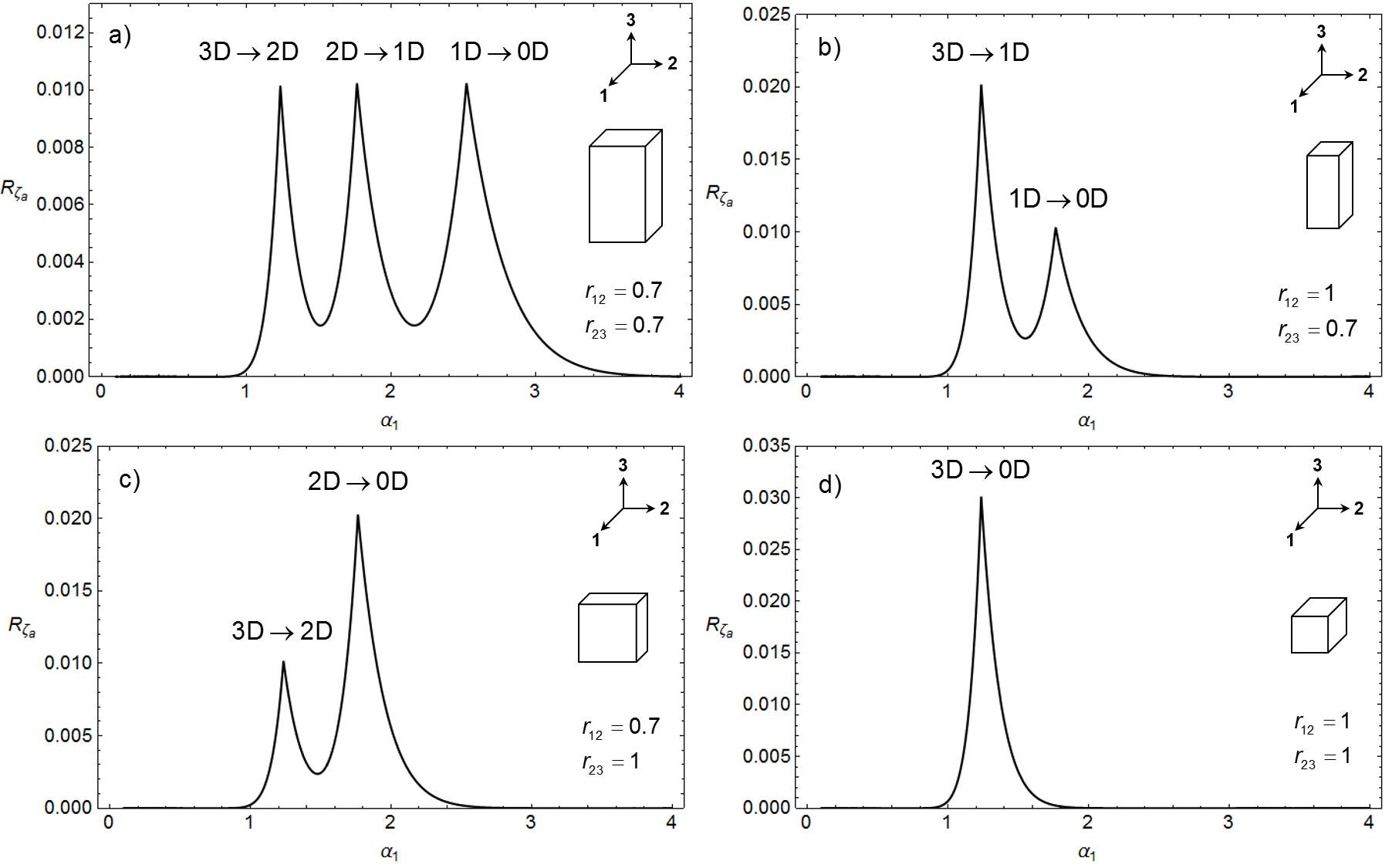}
\caption{Relative errors of analytical expressions of SPPF for four various aspect ratio sets. Peaks correspond to the relevant dimensional transition points. a) Fully anisometric domain with aspect ratios $r_{12}=r_{23}=0.7$. b) Semi-anisometric domain with aspect ratios $r_{12}=1$ and $r_{23}=0.7$. c) Semi-anisometric domain with aspect ratios $r_{12}=0.7$ and $r_{23}=1$. d) Isometric domain with aspect ratios $r_{12}=r_{23}=1$.}
\label{fig:pic2}
\end{figure}

Peaks in Figure 2 correspond to the values $\alpha^{*}$, $\alpha^{*}/r_{12}$ and $\alpha^{*}/r_{12}r_{23}$ for transitions from 3D-2D, 2D-1D and 1D-0D respectively. As it is seen from Figure 2a, each dimensional transition causes an error of $1\%$ even at the transition points. If the sizes in two different directions are the same, dimensional transitions in these directions superpose and it causes $2\%$ error as it is seen in Figures 2b and 2c. Similarly, for an isometric domain, superposition of three different dimensional transitions cause $3\%$ error, Figure 2d.

\section{Dimensional Transitions in Thermodynamic Properties}
Free energy of an ideal MB gas consisting of $N$ particles is simply given by $F=-N k_{b} T[\ln(\zeta/N)+1]$ \cite{kittel,pathbook}. Dimensionless free energy can be defined as $\widetilde{F}=F/Nk_BT$. Then, we can obtain analytical expressions for dimensionless free energies of a MB gas for 3D, 2D, 1D and 0D cases by considering dimensional transitions
\begin{equation}
\begin{split}
& \widetilde{F}_{3D}=-ln\left(\frac{\pi^{3/2}}{8N\alpha_{1}\alpha_{2}\alpha_{3}}\right)-1+ln\left(q_1 +1\right)+ln\left(q_2 +1\right)+ln\left(q_3 +1\right) \\
& \widetilde{F}_{2D}=-ln\left(\frac{\pi}{4N\alpha_{2}\alpha_{3}}\right)-1+\alpha_{1}^{2}+ln\left(q_2 +1\right)+ln\left(q_3 +1\right), \quad \{\alpha_1\}>\alpha^{*} \\
& \widetilde{F}_{1D}=-ln\left(\frac{\sqrt{\pi}}{2N\alpha_{3}}\right)-1+\alpha_{1}^{2} +\alpha_{2}^{2}+ln\left(q_3 +1\right), \quad \{\alpha_1,\alpha_2\}>\alpha^{*} \\
& \widetilde{F}_{0D}=-ln\left(\frac{1}{N}\right)-1+\alpha_{1}^{2} +\alpha_{2}^{2} + \alpha_{3}^{2}, \quad \{\alpha_1,\alpha_2,\alpha_3\}>\alpha^{*}
\end{split}
\end{equation}
respectively, where $q_n=\alpha_n/\left(\sqrt{\pi}-\alpha_n\right)$ with $n=\left\{1,2,3\right\}$ as direction index. It is clear that $\alpha_n$ is always smaller than $\sqrt{\pi}$ since $q_n$ is necessary only for the case of $\alpha_n<\alpha^{*}$. Note that, we can use MB statistics if and only if $1<<N<<1/\alpha_1 \alpha_2 \alpha_3$ condition holds. This condition holds even if one of the confinement parameters $\alpha_n$ is smaller enough than unity. Because in that condition the system contains enough particles to do statistics $(N>>1)$ and interparticular distance is much greater than $L_c$ to use MB statistics. Since 0D case does not satisfy this condition, one could not use MB statistics for 0D conditions. Nevertheless, in this article, we gave also 0D expressions of thermodynamic properties, just to give a mathematical intuition. Therefore, it is worth to note that thermodynamic expressions given for 0D do not represent a meaningful physical situation since the condition of $N<<1/\alpha_1 \alpha_2 \alpha_3$ leads $N<1$ for 0D condition.

From free energies, we can derive all other thermodynamic quantities such as entropy, internal energy, chemical potential, heat capacity at constant sizes and pressure as
\begin{subequations}
\begin{align*}
S=-&\left(\frac{\partial F}{\partial T}\right)_{N,V} \quad\quad U=F+TS \quad\quad \mu=\left(\frac{\partial F}{\partial N}\right)_{T,V} \\
&C=\left(\frac{\partial U}{\partial T}\right)_{L_1,L_2,L_3} \quad\quad P_{nn}=\frac{N}{V}k_B T\frac{L_n}{\zeta}\left(\frac{\partial \zeta}{\partial L_n}\right)_T
\end{align*}
\end{subequations}

Thermodynamic properties can be written in their dimensionless forms as $\widetilde{S}=S/Nk_B$, $\widetilde{U}=U/Nk_BT$, $\widetilde{\mu}=\mu/Nk_BT$, $\widetilde{C}=C /Nk_B$ and $\widetilde{P}_{nn}=P_{nn}V/Nk_BT$ where $V=L_1 L_2 L_3$. Even though pressure is a scalar quantity in macro scale, it becomes a tensorial quantity with zero off-diagonal elements for anisometric domains. In other words, $P_{11}\neq P_{22}\neq P_{33}$ for an anisometric rectangular domain where $L_{1}\neq L_{2}\neq L_{3}$.

Dimensionless entropy, internal energy, chemical potential, heat capacity and pressure expressions for MB gases are derived by considering dimensional transitions and given in Table 1.

\begin{table}[H]
\caption{Thermodynamic properties of Maxwell-Boltzmann gases in various dimensional conditions}
\centering
\renewcommand{\arraystretch}{1.5}
\begin{tabular}{l}
\hline
Entropy \\[0.5ex]
\hline
$\begin{array} {lcl} \widetilde{S}_{3D}=\frac{5}{2}+ln\left(\frac{\pi^{3/2}}{8N\alpha_{1}\alpha_{2}\alpha_{3}}\right)-ln\left(q_1 +1\right)-ln\left(q_2 +1\right)-ln\left(q_3 +1\right)+\frac{1}{2}\left(q_1 + q_2 + q_3\right) \end{array}$\\
$\begin{array} {r@{}l@{}} \widetilde{S}_{2D}=2+ln\left(\frac{\pi}{4N\alpha_{2}\alpha_{3}}\right)-ln\left(q_2 +1\right)-ln\left(q_3 +1\right)+\frac{1}{2}\left(q_2+q_3\right), \quad \{\alpha_1\}>\alpha^{*} \end{array}$\\
$\begin{array} {r@{}l@{}} \widetilde{S}_{1D}=\frac{3}{2}+ln\left(\frac{\sqrt{\pi}}{2N\alpha_3}\right)-ln\left(q_3 +1\right)+\frac{1}{2}(q_3), \quad \{\alpha_1,\alpha_2\}>\alpha^{*} \end{array}$\\
$\begin{array} {c} \widetilde{S}_{0D}=1+ln\left(\frac{1}{N}\right), \quad \{\alpha_1,\alpha_2,\alpha_3\}>\alpha^{*} \end{array}$\\
\hline
Internal energy \\[0.5ex]
\hline
$\begin{array} {lcl} \widetilde{U}_{3D}=\frac{3}{2}+ \frac{1}{2}\left(q_1 + q_2 + q_3\right) \end{array}$\\ 
$\begin{array} {r@{}l@{}} \widetilde{U}_{2D}=1+\alpha_{1}^2 + \frac{1}{2}\left(q_2 + q_3\right), \quad \{\alpha_1\}>\alpha^{*} \end{array}$\\ 
$\begin{array} {r@{}l@{}} \widetilde{U}_{1D}=\frac{1}{2}+\alpha_{1}^2 +\alpha_{2}^2 + \frac{1}{2}(q_3), \quad \{\alpha_1,\alpha_2\}>\alpha^{*} \end{array}$\\
$\begin{array} {c} \widetilde{U}_{0D}=\alpha_{1}^2 +\alpha_{2}^2 + \alpha_{3}^2, \quad \{\alpha_1,\alpha_2,\alpha_3\}>\alpha^{*} \end{array}$\\ 
\hline
Chemical potential \\[0.5ex]
\hline
$\begin{array} {lcl} \widetilde{\mu}_{3D}=-ln\left(\frac{\pi^{3/2}}{8N\alpha_{1}\alpha_{2}\alpha_{3}}\right)+ln\left(q_1 +1\right)+ln\left(q_2 +1\right)+ln\left(q_3 +1\right) \end{array}$\\
$\begin{array} {r@{}l@{}} \widetilde{\mu}_{2D}=-ln\left(\frac{\pi}{4N\alpha_{2}\alpha_{3}}\right)+\alpha_{1}^{2}+ln\left(q_2 +1\right)+ln\left(q_3 +1\right), \quad \{\alpha_1\}>\alpha^{*} \end{array}$\\
$\begin{array} {r@{}l@{}} \widetilde{\mu}_{1D}=-ln\left(\frac{\sqrt{\pi}}{2N\alpha_{3}}\right)+\alpha_{1}^{2} +\alpha_{2}^{2}+ln\left(q_3 +1\right), \quad \{\alpha_1,\alpha_2\}>\alpha^{*} \end{array}$\\
$\begin{array} {c} \widetilde{\mu}_{0D}=-ln\left(\frac{1}{N}\right)+\alpha_{1}^{2} +\alpha_{2}^{2} + \alpha_{3}^{2}, \quad \{\alpha_1,\alpha_2,\alpha_3\}>\alpha^{*} \end{array}$\\
\hline
Heat capacity at constant sizes $\left\{L_1,L_2,L_3\right\}$ \\[0.5ex]
\hline
$\begin{array} {lcl} \widetilde{C}_{3D}=\frac{3}{2}+\frac{1}{4}\left[q_1 (1-q_1) + q_2 (1-q_2) + q_3 (1-q_3)\right] \end{array}$\\
$\begin{array} {r@{}l@{}} \widetilde{C}_{2D}=1+\frac{1}{4}\left[q_2 (1-q_2) + q_3 (1-q_3)\right], \quad \{\alpha_1\}>\alpha^{*} \end{array}$\\
$\begin{array} {r@{}l@{}} \widetilde{C}_{1D}=\frac{1}{2}+\frac{1}{4}\left[q_3 (1-q_3)\right], \quad \{\alpha_1,\alpha_2\}>\alpha^{*} \end{array}$\\
$\begin{array} {c} \widetilde{C}_{0D}=0, \quad \{\alpha_1,\alpha_2,\alpha_3\}>\alpha^{*} \end{array}$\\
\hline
Pressure \\[0.5ex]
\hline
$\begin{array} {lcl} \widetilde{P}_{11}^{3D}=(q_1 +1), \quad \widetilde{P}_{22}^{3D}=(q_2 +1), \quad \widetilde{P}_{33}^{3D}=(q_3 +1) \end{array}$\\
$\begin{array} {r@{}l@{}} \widetilde{P}_{11}^{2D}=2\alpha_1^2, \quad \widetilde{P}_{22}^{2D}=(q_2 +1), \quad \widetilde{P}_{33}^{2D}=(q_3 +1), \quad \{\alpha_1\}>\alpha^{*} \end{array}$\\
$\begin{array} {r@{}l@{}} \widetilde{P}_{11}^{1D}=2\alpha_1^2, \quad \widetilde{P}_{22}^{1D}=2\alpha_2^2, \quad \widetilde{P}_{33}^{1D}=(q_3 +1), \quad \{\alpha_1,\alpha_2\}>\alpha^{*} \end{array}$\\
$\begin{array} {c} \widetilde{P}_{11}^{0D}=2\alpha_1^2, \quad \widetilde{P}_{22}^{0D}=2\alpha_2^2, \quad \widetilde{P}_{33}^{0D}=2\alpha_3^2, \quad \{\alpha_1,\alpha_2,\alpha_3\}>\alpha^{*} \end{array}$\\ 
\end{tabular}
\label{label 1}
\end{table}

It is easy to check that, all expressions derived in this article go to their classical expressions when confinement parameter $\alpha_n$'s go to zero. So they are valid not only for nano systems but also for meso and macro systems. They are generic in that sense, within the error limit at dimensional transition points.

Note for heat capacity that, instead of referring to constant volume, we refer here constant sizes which means size of each direction $L_1$, $L_2$ and $L_3$ are constant. Otherwise, combination of different $L_n$ values can also correspond to the same volume, although they correspond different heat capacity values, different from that in macro scale.

Unlike the scalar behavior of pressure in macroscopic scale, quantum confinement in one direction brings a positive contribution to the pressure in that direction which make the pressure a tensorial quantity in nanoscale. Besides, when confinement of a particular direction is larger than $\alpha^{*}$, pressure at that direction grows quadratically.

\section{Conclusion}

By introducing dimensional transitions of thermodynamic properties for MB gases, it's shown that it is possible to obtain analytical expressions within a negligible error to represent the exact forms of thermodynamic properties based on triple infinite summations. In this regard, analytical results are obtained for SPPF, free energy, entropy, internal energy, chemical potential, heat capacity and pressure, for 3D, 2D, 1D and 0D cases. It is seen that thermodynamic quantities have strict size dependency which cause dimensional contraction in free momentum space for nano systems.

Critical value of confinement parameter is found as $\alpha^{*}=1.23547...$. When the confinement parameter in a particular direction becomes greater than critical value of transition, momentum component in that direction can only take its ground-state value. Then, degree of freedom of momentum space decreases one unit. Therefore, depending on the values of confinement parameter in each direction, degree of freedom takes the values of $3,2,1$ and $0$. By considering this dimensional transition model, it is possible to express the thermodynamic properties analytically with $1\%$ error for any confined system.

As it is expected, classical expressions can be recovered in case of the absence of confinement, $(\alpha_n\rightarrow 0)$. Dimensional transitions in ideal Fermi and Bose gases are under consideration. Thermodynamic properties of strongly confined ideal Fermi and Bose gases in nano domains will then be analytically obtained for any scale.

\bibliography{wfrefc}{}
\bibliographystyle{unsrt}
\end{document}